\documentclass{article}
\usepackage{amssymb}
\usepackage{latexsym}
\usepackage{bez123}
\usepackage{multiply}

\title{Bethe ansatz and fluctuations in SU(3) Yang-Mills operators}
\author{Lisa Freyhult\footnote{lisa.freyhult@teorfys.uu.se}\\ Department of Theoretical Physics, Uppsala University\\P.O. Box 803, S-75108, Uppsala, Sweden}
\date{}

\begin{document}
\maketitle

\begin{abstract}
We consider the scalar operators corresponding to semiclassical string states in $AdS_5\times S^5$ with the three angular momenta in $S^5$ non-trivial. The string states recieve quantum corrections and we study the corresponding process on the gauge theory side. The anomalous dimension of the scalar operators is computed using the Bethe ansatz and we find the correction that corresponds to the energy of the quantized string. We restrict for simplicity to the case where two of the angular momenta in $S^5$ are equal.
\end{abstract}
\section{Introduction}
The AdS/CFT correspondence predicts the string theory in $AdS_5\times S^5$ to be dual to $N=4$ supersymmetric Yang-Mills theory \cite{Maldacena:1998re,Gubser:1998bc,Witten:1998qj}. One test of this conjecture involves computing the energy of semiclassical string states, this should then be compared with the scaling dimensions of local operators on the gauge theory side. Several examples of string solutions can be found in \cite{Gubser:2002tv,Frolov:2002av,Russo:2002sr,Minahan:2002rc,Tseytlin:2002ny,Frolov:2003qc,Frolov:2003tu,Frolov:2003xy,Arutyunov:2003uj,Khan:2003sm,Arutyunov:2003rg,Arutyunov:2003za,Mikhailov:2003gq,Stefanski:2003qr,Arutyunov:2004xy,Dimov:2004xi,Smedback:2004yn} and their gauge theory duals in \cite{Minahan:2002ve,Beisert:2002ff,Belitsky:2003ys,Beisert:2003xu,Beisert:2003jj,Beisert:2003yb,Beisert:2003ea,Gorsky:2003nq,Engquist:2003rn,Arutyunov:2003rg,Kristjansen:2004ei,Lubcke:2004dg,Engquist:2004bx,Arutyunov:2004xy,Kazakov:2004qf}. One of these examples is the recently found solution describing a circular string rotating in two orthogonal planes in the $S^5$ part of space with equal angular momenta, the center of mass orbiting around another circle in $S^5$ \cite{Frolov:2003qc} \cite{Frolov:2003tu}.  The corresponding gauge theory computation is discussed in \cite{Engquist:2003rn}. This type of solution simplifies thanks to the fact that two of the angular momenta are equal. The energy of the classical string solution can be expanded in $\lambda/L^2$. For the string $\sqrt\lambda$ is proportional to the string tension and here $L$ is the sum of the three angular momenta of the rotating string. We will consider the limit where $L$ is large. It is then possible quantize the solutions and hence find quantum corrections to their energy. The energy, including quantum corrections, will be of the form
\begin{equation}
E=E_0(J,J')+\frac{\lambda}{L^2}E_1(J,J')+\left(\frac{\lambda}{L^2}\right)^2E_2(J,J')+...
\end{equation}
where $J$ and $J'$ are the two different angular momenta. It will be possible to expand $E_1$, $E_2$ etc. in $1/L$ where the first term in the expansion corresponds to the classical result. On the string side the first term in the expansion is found introducing fluctuations in the string action and considering each field split into a fast and a slowly varying part. Solutions to the effective action are then found and their energies computed.

On the gauge side the perturbative expansion of the dimension of the operator that corresponds to the string solution in question, is the dual of the energy of the classical string. Computing the anomalous dimension involves the difficulty that in general scalar operators mix under renormalization. This problem is solved by identifying the matrix of anomalous dimensions with the Hamiltonian of a spin chain as described in \cite{Minahan:2002ve}. The spin chain is an integrable system and can be solved using the Bethe ansatz \cite{Bethe:1931hc}\cite{Faddeev:1996iy}. In \cite{Minahan:2002ve} the $SO(6)$ spin chain, associated to the planar one-loop dilatation operator for scalar operators, was discussed. This has been generalized and additional types of spin chains have been introduced \cite{Beisert:2003tq,Beisert:2003jj,Beisert:2003yb,Beisert:2003ys,Serban:2004jf,Beisert:2004hm}.

In \cite{Engquist:2003rn} the gauge dual of the classical string with spins $(J,J',J')$ was found, here we will consider the corrections that corresponds to the dual of the quantized string. Previously similar methods have been used to find the corrections that correspond to the dual of a string where one of the angular momenta is zero \cite{Beisert:2003xu}.

In section 2 we will briefly describe the gauge theory computations and the Bethe ansatz used to diagonalize the anomalous dimension matrix. In section 3 the integral version of the relevant Bethe equations, before introducing fluctuations, are reviewed. In section 4 and 5 these equations will be used and modified in order to take fluctuations to the solutions into account. At the end of section 5 we arrive at an expression for the anomalous dimension with the fluctuations taken into account. 
\section{Gauge theory computations}
What we will study here will be the duals of the semiclassical solutions to the string sigma model describing rotating strings in the $S^5$ sector of $AdS_5\times S^5$. It is assumed that the angular momenta of the strings is large in order to allow for an expansion in the parameter $1/L$. In \cite{Frolov:2003qc} it was argued that these string states are dual to the SYM operators of the form
\begin{equation}
O=\mbox{Tr}X^{J_1}Y^{J_2}Z^{J_3}+...
\end{equation}
where $X$, $Y$ and $Z$ can be expressed in terms of the six scalar fields in $N=4$ supersymmetric Yang-Mills theory as follows $X=\Phi_1+i\Phi_2$, $Y=\Phi_3+i\Phi_4$, $Z=\Phi_5+i\Phi_6$. The dots in the operator stand for permutations of the different constituent fields. The bare dimension of this operator is $\Delta_0=J_1+J_2+J_3$ and computing the one-loop correction to this it is possible to show that the matrix of anomalous dimension is equivalent to the Hamiltonian of an integrable spin chain with $SO(6)$ symmetry \cite{Minahan:2002ve}.  This matrix is in general difficult to diagonalize however associating it to a spin chain it is possible to use the methods of the Bethe ansatz to perform the diagonalization. As was found in \cite{Minahan:2002ve} the problem can be reduced to solving the Bethe equations for the Bethe roots. The Bethe equations are
\begin{eqnarray}
\nonumber\left(\frac{u_{1,i}+i/2}{u_{1,i}-i/2}\right)^L&=&\prod_{j\neq i}^{n_1}\frac{u_{1,i}-u_{1,j}+i}{u_{1,i}-u_{1,j}-i}\prod_j^{n_2}\frac{u_{1,i}-u_{2,j}-i/2}{u_{1,i}-u_{2,j}+i/2}\prod_j^{n_3}\frac{u_{1,i}-u_{3,j}-i/2}{u_{1,i}-u_{3,j}+i/2}\\
\nonumber1&=&\prod_{j\neq i}^{n_2}\frac{u_{2,i}-u_{2,j}+i}{u_{2,i}-u_{2,j}-i}\prod_j^{n_1}\frac{u_{2,i}-u_{1,j}-i/2}{u_{2,i}-u_{1,j}+i/2}\\
1&=&\prod_{j\neq i}^{n_3}\frac{u_{3,i}-u_{3,j}+i}{u_{3,i}-u_{3,j}-i}\prod_j^{n_1}\frac{u_{3,i}-u_{1,j}-i/2}{u_{3,i}-u_{1,j}+i/2}
\end{eqnarray}
Here $n_1$, $n_2$ and $n_3$ is the number of Bethe roots of each type. The different types of Bethe roots are associated to the simple roots of $SO(6)$. It can also be shown that the anomalous dimension is given by
\begin{equation}\label{anomdimsum}
\gamma=\frac{\lambda}{8\pi^2}\sum_i^{n_1}\frac{1}{(u_{1,i})^2+1/4}.
\end{equation}
In the termodynamic limit, i.e. when the number of sites in the spin chain becomes large, it is possible to rewrite the Bethe equations as integral equations which makes them considerably much easier to solve. In the following we will consider the specific case where two of the angular momenta are equal. There will be two $SO(6)$ representations, one for $J'\leq J$ and one for $J'\geq J$, where $J$ and $J'$ are the two possible angular momenta.
In the representation where $J'\leq J$ we will have 
\begin{equation}
n_2=n_1/2, n_3=0\quad\mbox{and}\quad J_1=J, J_2=J_3=J',n_2=J', n_1=L-J.
\end{equation}
In the $J'\geq J$ representation we have
\begin{equation}
n_1=L/2+n_2/2, n_3=0\quad\mbox{and}\quad J_1=J_2=J', J_3=J,J'=n_1-n_2, J=n_2.
\end{equation}
Each representation will give a different distribution of the roots in the complex plane and therefore lead to qualitatively different equations to solve. However it is possible to show that the two cases can be related by analytic continuation \cite{Engquist:2003rn}. We also expect that the anomalous dimension should be independent of the representation chosen. Previously it has been verified that the first $1/L$ correction to the dimension of the operators exactly coincide with the corresponding string theory result \cite{Engquist:2003rn}. On the string side there is also a result for the next order correction corresponding to quantizing the string. This is done by shifting the fields in the action and then integrating out the higher momentum part of the fields. This way an effective action is obtained and the corrections to the spectrum is computed \cite{Frolov:2003qc} \cite{Frolov:2003tu}. On the gauge side the corresponding corrections comes from moving Bethe roots around in the complex plane. This has previously been described in \cite{Beisert:2003xu} for the case with one of the angular momenta equal to zero and the other two equal. This case corresponds to $J_3=0$ in the representation $J'\geq J$ and to $J_1=0$ in the other representation. What we will consider in the following is the the correction coming from fluctuations in the Bethe roots on the gauge side for the more general configuration of angular momenta described above. We will do the calculation in the representation where $J'\leq J$ since the problem simplifies remarkably there.

\section{The integral Bethe equations in the representation $J'\leq J$}
The Bethe equations are converted to integral equations by taking the logarithm of the equations and rescaling the roots, $u=qL$. In the large $L$ limit the sums can then be converted into integrals. The result in our case is 
\begin{eqnarray}\label{Bethe1}
\frac{1}{q}-2\pi m&=&\alpha-\hspace{-0.4cm}\int_{C_+}dq'\frac{\sigma(q')}{q-q'}+\alpha\int_{C_+}dq'\frac{\sigma(q')}{q+q'}-\beta\int_{C}dq'\frac{\rho(q')}{q-q'}\\\label{Bethe2}
0&=&2\beta-\hspace{-0.4cm}\int_{C}\frac{\rho(q')}{q-q'}-\frac{\alpha}{2}\int_{C_+}dq'\frac{\sigma(q')}{q-q'}-\frac{\alpha}{2}\int_{C_+}dq'\frac{\sigma(q')}{q+q'}
\end{eqnarray}
Here we have introduced $\alpha=n_1/L$ and $\beta=n_2/L$. We have also defined the densities of roots as
\begin{equation}\label{densities}
\sigma(q)=\frac{2}{\alpha L}\sum_i^{n_1}\delta(q-q_i)\quad \rho(q)=\frac{2}{\beta L}\sum_i^{n_2}\delta(q-q_i).
\end{equation}
where $\sigma(q)$ is the density of the first type of roots and $\rho(q)$ the density of the second type roots. The integer $m$ in eq. (\ref{Bethe1}) comes from the branch of the logarithm. For this representation the roots will be distributed as shown in figure \ref{Figure1}.
\setlength{\unitlength}{1cm}
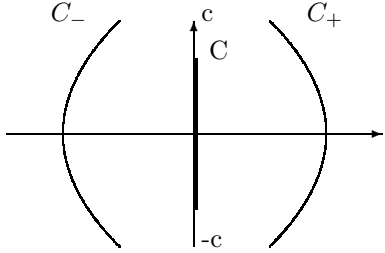
\begin{figure}
\begin{picture}(5,5)
\put(0,2.5){\vector(1,0){5}}
\put(2.5,1){\vector(0,1){3}}
\put(2.53,1.5){\line(0,1){2}}
\qbezier(1.5,1)(0,2.5)(1.5,4)
\qbezier(3.5,1)(5,2.5)(3.5,4)
\put(0.6,4){$C_-$}
\put(4,4){$C_+$}
\put(2.6,1){-c}
\put(2.6,4){c}
\put(2.7,3.5){C}
\end{picture}
\vspace{-1cm}
\caption{\label{Figure1}The distribution of roots in the complex plane.}
\end{figure}
The densities will be normalized such that
\begin{equation}\label{normalization}
\int_{C_+}dq\sigma(q)=1\quad \int_Cdq\rho(q)=1.
\end{equation}
The anomalous dimension in (\ref{anomdimsum}) can be rewritten as
\begin{equation}\label{anomdim}
\gamma=\frac{\lambda\alpha}{8\pi^2L}\int_{C_+}dq\frac{\sigma(q)}{q^2}.
\end{equation}

Solving eq. (\ref{Bethe2}) we find that
\begin{equation}\label{rho}
\rho(iq)=\frac{\alpha}{2\beta\pi}\int_{C_+}dq'\frac{\sigma(q')q'}{(q')^2+q^2}\frac{\sqrt{c^2-q^2}}{\sqrt{c^2+q'^2}}
\end{equation}
Here $c$ is the endpoint of the distribution of second type roots along the imaginary axis as indicated in figure \ref{Figure1}.
Substituting this into the normalization condition (\ref{normalization}) we find the condition
\begin{equation}
\frac{\alpha}{2\beta}-\frac{\alpha}{2\beta}\int_{C_+}\frac{\sigma(q)}{\sqrt{c^2+q^2}}=1
\end{equation}
In the limit where $c\to\infty$ this means that $\beta=\alpha/2$. This is called the half-filling condition. At half-filling using eq. (\ref{rho}) it is possible to derive
\begin{equation}\label{helprel}
q\int_{-\infty}^\infty dq'\frac{\rho(iq')}{q^2+q'^2}=\int_{C_+}dq'\frac{\sigma(q')}{q+q'}
\end{equation}
Note that this relation holds for $q>0$, which is all we will need at the moment.
Using this in (\ref{Bethe1}) the following equation is obtained
\begin{equation}\label{potential}
\frac{2}{\alpha}\left(\frac{1}{q}-2\pi m\right)=2-\hspace{-0.4cm}\int dq'\frac{\sigma(q')}{q-q'}+\int_{C_+}\frac{\sigma(q')}{q+q'}
\end{equation}
In order to simplify the notation in the following we will introduce the resolvent
\begin{equation}\label{resolvent}
W(q)=\int_{C_+}dq'\frac{\sigma(q')}{q-q'}
\end{equation}

In order to find the corrections from fluctuations we will move Bethe roots around in the complex plane. 
Here we consider spinless fluctuations and therefore the roots can not be moved such that they change the representation. For this reason we have to move existing roots, new roots would change the representation. The simplest possible fluctuation is to move two roots symmetrically to the real axis. If the roots are of the first type their positions will directly contribute as can be seen from (\ref{anomdimsum}). There will also be an interaction between the moved roots and the roots left in $C_+$, $C_-$ and $C$. This will modify the densities of roots. The modified densities will then affect the position of the fluctuations and so on, this however will be higher order corrections in the expansion parameter $1/L$. Here we will consider corrections to order $1/L^2$ but not higher and we can therefore safely ignore those higher order effects.
If the roots are of the second type, they can only affect the anomalous dimension through the backreaction. However we will show that two such roots can only be taken to infinity.
 
\section{The positions of the moved roots}
\subsection{The positions of the fluctuations corresponding to the first type of roots}
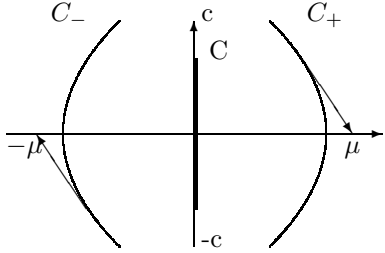
\begin{figure}
\begin{picture}(5,5)
\put(0,2.5){\vector(1,0){5}}
\put(2.5,1){\vector(0,1){3}}
\put(2.53,1.5){\line(0,1){2}}
\qbezier(1.5,1)(0,2.5)(1.5,4)
\qbezier(3.5,1)(5,2.5)(3.5,4)
\put(0.6,4){$C_-$}
\put(4,4){$C_+$}
\put(2.6,1){-c}
\put(2.6,4){c}
\put(2.7,3.5){C}
\put(3.94,3.5){\vector(2,-3){0.66}}
\put(4.5,2.25){$\mu$}
\put(1.07,1.5){\vector(-2,3){0.66}}
\put(0,2.25){$-\mu$}
\end{picture}
\vspace{-1cm}
\caption{\label{Figure2}Two roots of the first type is moved to the real axis. $\mu$ could be located anywhere on the real line except at the intersection with $C_+$.}
\end{figure}

We start by moving one root from $C_+$ and one from $C_-$ symmetrically to the real axis, as shown in figure \ref{Figure2}. This will not change the representation as we will se in the following. Here we assume that the fluctuations will not be located at the point where $C_+$ and $C_-$ crosses the real axis. The motivation for that can be found from the Bethe equations; they show that roots of the same type repulse each other. Hence if we try to move one of the roots from one point in $C_+$ to another the density will change. The roots will then interact and we will again reach the equilibrium position that we started with. Therefore moving roots within $C_\pm$ will not change the distribution of the roots and hence it will not give any contribution to the anomalous dimension.

We use the first Bethe equation (\ref{Bethe1}) and (\ref{helprel}) to determine the position of the roots, $\mu$. 
\begin{equation}
2W(\mu)-W(-\mu)=\frac{1}{\mu}-2\pi n
\end{equation}
where $W(q)$ is the resolvent given in (\ref{resolvent}) and $n$ is a positive integer. Since we have assumed that the moved root does not lie on $C_+$ we have that $n\neq m$. In the following we will assume that the position of the root on the real axis, $\mu$, is positive.
The resolvent can be written as $W(q)=W_p(q)+w(q)$ where
\begin{equation}
W_p(q)=\frac{1}{3}(2U'(q)+U'(-q))
\end{equation}
for positive $q$.
The potential is given by
\begin{equation}
U'(q)=\frac{1}{q}-2\pi m
\end{equation}
as seen from eq. (\ref{potential}).
This means
\begin{equation}
W_p(\mu)=\frac{2}{3\alpha\mu}-\frac{4\pi m}{\alpha}
\end{equation}
Using that $W(q)=W_p(q)+w(q)$ the equation for the fluctuations reduces to
\begin{equation}\label{eomfluct}
2w(\mu)-w(-\mu)=4\pi(m-n)
\end{equation}
It is possible to show that $w(q)$ satisfy
\begin{equation}\label{rorig}
w(q)^2-w(q)w(-q)+w(-q)^2=r(q)
\end{equation}
where $r(q)=\left(\frac{4\pi m}{\alpha}\right)^2+\frac{4}{3\alpha^2q^2}$, see \cite{Engquist:2003rn} for details. We also have the following relation
\begin{equation}\label{sorig}
w(q)^3-r(q)w(q)=s(q)
\end{equation}
where $s(q)=\frac{16}{27\alpha^2 q^3}+2\left(\frac{4\pi m}{\alpha}\right)^2\left(1-\frac{2}{3\alpha}\right)\frac{1}{q}$.
(\ref{eomfluct}) together with (\ref{sorig}) gives that
\begin{eqnarray}\label{w(mu)}
&&\hspace{-1cm}\nonumber w(\mu)=\frac{2\pi}{\alpha}(m-1-p)\pm\frac{2}{3\alpha\mu}\sqrt{3\mu^2\pi^2(-p^2+(2m-1)p+3m^2+2m-1)+1}\\
&&\hspace{-1cm}w(-\mu)=\pm\frac{4}{3\alpha\mu}\sqrt{3\mu^2\pi^2(-p^2+(2m-1)p+3m^2+2m-1)+1}
\end{eqnarray}
Here we have made the shift $n=m(p+1)$ in order to simplify the formulas in the following and in order to be able to identify our result with the result on the string side \cite{Frolov:2003tu}. 
Using eq. (\ref{sorig}) we derive the position of the fluctuations
\begin{eqnarray}\label{positions}
\nonumber\mu =\pm\frac{\sqrt2}{4\pi m (p^2-1)\sqrt{p^2-4}}\times\hspace{6.5cm}\\
\sqrt{36\alpha-27\alpha^2-8+2p^4-6p^2\pm(8-9\alpha-2p^2)\sqrt{9\alpha^2-(4p^2+8)\alpha+4p^2}}
\end{eqnarray}
In the special cases where $p\neq\pm1$ and $p\neq\pm2$ the solutions are
\begin{eqnarray}
&&\mu=\pm\frac{\sqrt{9-12\alpha}}{6\pi m(3\alpha-2)}\quad{p=\pm1, \alpha\neq 2/3}\\
&&\mu=\pm\frac{1}{3m\pi\sqrt{4-3\alpha}}\quad{p=\pm2}\\
&&\mu=0\hspace{2.5cm}{p=\pm1, \alpha=2/3}
\end{eqnarray}
The signs in the above expressions for $\mu$ should be chosen such that $\mu$ is always positive, as assumed earlier. At $\alpha=2/3$ the configuration of Bethe roots in the complex plane change as this corresponds to $J=J'$. The behaviour at that point seen above is an artifact of the Bethe ansatz.

Here we just note that in the limit where $\alpha=1$ this reduces to
\begin{equation}
\mu_{\alpha=1}=\frac{1}{2\pi}\frac{1}{\sqrt{p^2-4}}
\end{equation}
if we chose the branch corresponding to $m=1$.
This should be compared with the result in \cite{Beisert:2003xu} and is seen to agree. Note that in order to compare the results the integer $p$ has to be shifted by two.

The contribution to the anomalous dimension from the moved roots is
\begin{equation}
\gamma_1=\frac{2\lambda}{8\pi^2L^2\mu^2}
\end{equation}
which follows from (\ref{anomdimsum}).

\subsection{The positions of the fluctuations corresponding to the second type of roots}
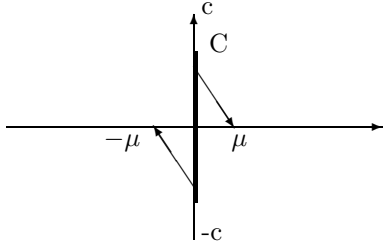
\begin{figure}
\begin{picture}(5,5)
\put(0,2.5){\vector(1,0){5}}
\put(2.5,1){\vector(0,1){3}}
\put(2.53,1.5){\line(0,1){2}}
\put(2.6,1){-c}
\put(2.6,4){c}
\put(2.7,3.5){C}
\put(2.5,3.3){\vector(2,-3){0.54}}
\put(3,2.25){$\mu$}
\put(2.5,1.7){\vector(-2,3){0.54}}
\put(1.3,2.25){$-\mu$}
\end{picture}
\vspace{-1cm}
\caption{\label{Figure3}Two roots of the second type is moved to the real axis.}
\end{figure}

We now consider moving a pair of second type roots symmetrically to the real axis, as shown in figure \ref{Figure3}. This will be the simplest case since by moving the roots symmetrically all the other roots in $C$ will stay on the imaginary axis.
From (\ref{Bethe2}) and (\ref{helprel}) it follows that the second type of roots satisfies the equation
\begin{equation}
0=W(\mu)-W(-\mu)
\end{equation}
Using this together with (\ref{rorig}) and (\ref{sorig}) we obtain the following equation.
\begin{equation}
-\frac{32\pi^2n^2}{\alpha^2\mu}=0
\end{equation}
This means that the moved roots are located at infinity. Roots at infinity do not affect the other roots and they do not contribute to the anomalous dimension. Moving roots to infinity is the same as completely removing them from the system. However removing roots means changing the representation which we do not want to do. 
\section{Contribution to $\gamma$ from the modified densities}
The moved roots will give $1/L$ corrections to the Bethe equations, this means that the densities could be affected. In the following we will determine the contribution to $\gamma$ from that effect.
The Bethe equations with two first type roots moved from $C_+$ and $C_-$ are
\begin{eqnarray}\label{Bethe1mod}
\hspace{-0.5cm}\frac{1}{q}-2\pi m\hspace{-0.26cm}&=&\hspace{-0.26cm}\alpha-\hspace{-0.4cm}\int_{C_+}dq'\frac{\sigma(q')}{q-q'}+\alpha\int_{C_+}dq'\frac{\sigma(q')}{q+q'}-\beta\int_{C}dq'\frac{\rho(q')}{q-q'}+\frac{2}{L}\frac{2q}{q^2-\mu^2}\\\label{Bethe2mod}
0\hspace{-0.1cm}&=&\hspace{-0.26cm}2\beta-\hspace{-0.4cm}\int_{C}\frac{\rho(q')}{q-q'}-\frac{\alpha}{2}\int_{C_+}dq'\frac{\sigma(q')}{q-q'}-\frac{\alpha}{2}\int_{C_+}dq'\frac{\sigma(q')}{q+q'}-\frac{1}{L}\frac{2q}{q^2-\mu^2}
\end{eqnarray}
Using the second Bethe equation (\ref{Bethe2mod}) to solve for $\rho(q)$ we find
\begin{equation}\label{rhomod}
\rho(iq)=-\frac{1}{2\pi^2\beta}\sqrt{c^2-q^2}-\hspace{-0.4cm}\int_c^c\frac{dq'}{q-q'}\frac{1}{\sqrt{c^2-q^2}}\left(\int_{C_+}dq''\frac{\sigma(q'')q'}{q'^2+q''^2}+\frac{1}{L}\frac{2q'}{q'^2+\mu^2}\right)
\end{equation}
Previously the limit $c\to\infty$ meant half-filling. Let us show that $c$ remains at infinity. 
If $c\to\infty$ eq. (\ref{rhomod}) reduces to
\begin{equation}\label{rhomodcinfty}
\rho(iq)=\frac{\alpha}{2\beta}\frac{1}{\pi}\int_{C_+}dq''\frac{\sigma(q'')}{q^2+q''^2}+\frac{1}{\pi\beta L}\frac{\mu}{q^2+\mu^2}
\end{equation}
Considering this together with the normalization conditions will give an extra condition. 
The normalization conditions for the distributions of densities are modified compared to the system without modifications from fluctuations. One root is moved from each of the distributions $C_\pm$ and hence 
\begin{eqnarray}\label{normalizationmod}
&&\nonumber\int_{C}\rho(q)dq=1\\
&&\int_{C_+}\sigma(q)dq=1-\frac{2}{\alpha L},
\end{eqnarray}
this follows directly from the definition of densities (\ref{densities}).
From eq. (\ref{rhomodcinfty}) above it follows that
\begin{equation}
\int_{-\infty}^\infty dq\rho(iq)=\frac{\alpha}{2\beta}\int_{C_+}dq\sigma(q)+\int_{-\infty}^{\infty}dq\frac{1}{\pi\beta L}\frac{\mu}{q^2+\mu^2}
\end{equation}
Using (\ref{normalizationmod}) we find
\begin{equation}
1=\frac{\alpha}{2\beta}\left(1-\frac{2}{\alpha L}\right)+\frac{1}{\beta L}\leftrightarrow\alpha=2\beta
\end{equation}
This is the half-filling condition as we set out to show.

From eq. (\ref{rhomodcinfty}) it is now possible to find the following relation
\begin{eqnarray}
q\int_{-\infty}^{\infty}dq'\frac{\rho(iq')}{q^2+q'2}=\int_{C_+}dq'\frac{\sigma(q')}{q+q'}+\frac{2}{\alpha L}\frac{1}{q+\mu}\quad q\in C_+
\end{eqnarray}
Note that we will consider the case where $q\in C_+$ since that is what we will need in the following.
Using this in eq. (\ref{Bethe1mod}) we obtain 
\begin{equation}\label{q>0}
\frac{2}{\alpha}\left(\frac{1}{q}-2\pi m\right)+\frac{1}{\beta L}\frac{1}{q+\mu}-\frac{8}{\alpha L}\frac{q}{q^2-\mu^2}=2-\hspace{-0.35cm}\int_{C_+}dq'\frac{\sigma(q')}{q-q'}+\int_{C_+} dq'\frac{\sigma(q')}{q+q'}
\end{equation}
for $q\in C_+$
Using the definition of the resolvent, (\ref{resolvent}), eq. (\ref{q>0}) can be rewritten as
\begin{eqnarray}
&&\nonumber W(q+i0)+W(q-i0)-W(-q)=U'(q)\\
&&\mbox{where}\quad U'(q)=\frac{2}{\alpha}\left(\frac{1}{q}-2\pi m\right)+\frac{1}{\beta L}\frac{1}{q+\mu}-\frac{8}{\alpha L}\frac{q}{q^2-\mu^2}
\end{eqnarray}
The contribution from the contour to the anomalous dimesion is
\begin{equation}
\gamma=-\frac{\lambda}{8\pi^2L}\alpha W'(0).
\end{equation}
Particular solution to the equation for the resolvent is 
\begin{eqnarray}
W_{p}(q)=\frac{1}{3}\left(2U'(q)+U'(-q)\right) 
\end{eqnarray}
This means that
\begin{equation}\label{particular}
W_p=-\frac{4\pi m}{\alpha}+\frac{2}{3\alpha q}-\frac{8}{3\alpha L}\frac{q}{q^2-\mu^2}+\frac{2}{3\alpha L}\left(\frac{2}{q+\mu}-\frac{1}{q-\mu}\right)
\end{equation}
The general solution is then
\begin{equation}
W(q)=W_p(q)+w(q)
\end{equation}
It is possible to determine the form of $w(q)$ using that $W(q)$ is regular at $q=0$ and $q=\pm\mu$. It is also possible to derive the behaviour of the resolvent as $q\to\infty$. Using the definition (\ref{resolvent}) and the normalization condition (\ref{normalizationmod}) we find that
\begin{equation}
W(q)=\frac{1}{q}\int_{C_+}dq'\sigma(q')=\frac{1}{q}\left(1-\frac{2}{\alpha L}\right)\quad\mbox{as $q\to\infty$}.
\end{equation}
This determines the form of $w(q)$ since it must be defined to cancel off singular terms from $W_r(q)$ and give the the correct asymptotic behaviour for $W(q)$.
\begin{eqnarray}\label{wasymptotic}
\nonumber w(q)&\to&\frac{4\pi m}{\alpha}+\left(1-\frac{2}{3\alpha}\right)\frac{1}{q}\quad q\to\infty\\
\nonumber w(q)&\to&-\frac{2}{3\alpha q}\hspace{2.6cm} q\to0\\
\nonumber w(q)&\to&\frac{2}{\alpha L}\frac{1}{q-\mu}+c_1\hspace{1.35cm} q\to\mu\\
w(q)&\to& c_2\hspace{3.2cm}q\to-\mu
\end{eqnarray}
$c_1$ and $c_2$ are constants to be determined later.
It is now possible to define an even function analogous to (\ref{rorig}).
\begin{equation}
r(q)=w(q)^2-w(-q)w(q)+w(-q)^2
\end{equation}
From the behavior  of the function $w(q)$ we find $r(q)$ in the different limits,
\begin{eqnarray}
\nonumber r(q)&\to&\left(\frac{4\pi m}{\alpha}\right)^2\hspace{2cm}q\to\infty\\
\nonumber&\to&\frac{4}{3\alpha^2q^2}\hspace{2.6cm}q\to0\\
\nonumber&\to&(2c_1-c_2)\frac{2}{\alpha L}\frac{1}{q-\mu}\hspace{0.6cm} q\to\mu\\
 &\to&\hspace{-0.3cm}-(2c_1-c_2)\frac{2}{\alpha L}\frac{1}{q+\mu}\hspace{0.7cm} q\to\mu.
\end{eqnarray}
This means that the only possible form of $r(q)$ is
\begin{equation}\label{r}
r(q)=\left(\frac{4\pi m}{\alpha}\right)^2+\frac{4}{3\alpha^2q^2}+(2c_1-c_2)\frac{2}{\alpha L}\frac{2\mu}{q^2-\mu^2}
\end{equation}
Multiplying $r(q)$ with $w(q)+w(-q)$ we find that
\begin{equation}\label{wrs}
w(q)^3-r(q)w(q)=-w(-q)^3+r(-q)w(-q)\equiv s(q)
\end{equation}
The behaviour of $s(q)$ can be determined from the above, eq. (\ref{wasymptotic}) and (\ref{r}),
\begin{eqnarray}
\nonumber s(q)&\to&2\left(\frac{4\pi m}{\alpha}\right)^2\left(1-\frac{2}{3\alpha}\right)\frac{1}{q}\hspace{3.5cm}q\to\infty\\
&\to&\nonumber \frac{16}{27\alpha^3q^3}\hspace{6cm}q\to0\\
&\to&\nonumber\left(c_1^2+c_1c_2-\left(\frac{4\pi m}{\alpha}\right)^2-\frac{4}{3\alpha^2\mu^2}\right)\frac{2}{\alpha L}\frac{1}{q-\mu}\quad q\to\mu\\
&\to&\left(2c_1c_2-c_2^2\right)\frac{2}{\alpha L}\frac{1}{q+\mu}\hspace{3.7cm}q\to-\mu
\end{eqnarray}
The function that satisfies this behaviour is
\begin{equation}\label{s}
s(q)=\frac{16}{27\alpha^3q^3}+2\left(\frac{4\pi m}{\alpha}\right)^2\left(1-\frac{2}{3\alpha}\right)\frac{1}{q}+\frac{2}{\alpha L}\frac{2\mu}{q^2-\mu^2}\frac{1}{q}C.
\end{equation}
Here we have introduced the constant $C$ which is given by 
\begin{eqnarray}
&&C=c_1^2+c_1c_2-\left(\frac{4\pi m}{\alpha}\right)^2-\frac{4}{3\alpha^2\mu^2}=2c_1c_2-c_2^2\\
&&=\frac{16\pi}{3\mu\alpha^2}\sqrt{3\mu^2\pi^2(3m^2+2m+2pm-(p+1)^2)+1}.
\end{eqnarray}
In order to find the above equality we have used that the constants are given by
\begin{equation}\label{constants}
c_1=w(\mu)\quad\quad c_2=w(-\mu),
\end{equation}
found in (\ref{w(mu)}). 
This is true only up to corrections of order $1/L$ but in fact that will be all that is needed.
$1/L$ corrections to the constants above will lead to corrections to the anomalous dimension of order $1/L^3$ and higher.

In order to find the anomalous dimension we solve eq. (\ref{s}) for $w(q)$, giving us the expression for the full resolvent. In fact we do not need the exact expression for the resolvent, it will be enough to have the first few terms in the expansion about $q=0$. 
\begin{equation}
w(q)=\frac{a_1}{q}+a_2+a_3q+a_4q^2+...
\end{equation}
Using (\ref{wrs}), (\ref{r}) and (\ref{s}) we find
\begin{eqnarray}\label{coeffs}
&&\nonumber a_1=-\frac{2}{3\alpha}\\
&&a_3=4\pi^2m^2+\frac{4\alpha\mu(2c_1-c_2)+3\alpha^2\mu^2(c_1^2+c_1c_2)-48\pi^2m^2\mu^2-4}{6\alpha\mu^2 L}
\end{eqnarray}
 From eq. (\ref{anomdim}), (\ref{particular}) and (\ref{coeffs}) and  we find
\begin{equation}
\gamma_2=-\frac{\lambda}{8\pi^2L}\alpha(W_p'(0)+w'(0))=-\frac{\lambda\alpha}{8\pi^2L}\left(\frac{2}{\alpha\mu^2 L}+a_3\right)
\end{equation}
Using the expressions for the constants $c_1$ and $c_2$, from (\ref{constants}) and (\ref{w(mu)}) we find that the full anomalous dimension, including the contribution from the moved roots is 
\begin{equation}
\gamma=\gamma_1+\gamma_2=\frac{\lambda\alpha m^2}{2L}+\frac{\lambda}{L^2}\frac{1}{6\mu\pi}\left(2p+2p\sqrt{3\mu^2\pi^2(4-p^2)+1}+6\mu\pi(m^2-1)\right)
\end{equation}
In order to simplify the expression we chose the branch corresponding to $m=1$, this corresponds to the choice of parameters in \cite{Frolov:2003tu}.
Using the positions of the moved roots, from (\ref{positions}), we then find 
\begin{eqnarray}
&&\hspace{-0.5cm}\nonumber\gamma=\left(\frac{\lambda}{2L}+\frac{\lambda}{L^2}p\sqrt{p^2-4}\right)+\left(\frac{\lambda}{2L}+2\frac{\lambda}{L^2}p\sqrt{p^2-4}\right)(\alpha-1)\\
&&\nonumber\hspace{-0.5cm}+\frac{2\lambda}{L^2}p(p^2-2)\sqrt{p^2-4}(\alpha-1)^2+\frac{2\lambda}{L^2}(2p^4-9p^2+9)p\sqrt{p^2-4}(\alpha-1)^3\\
&&\hspace{-0.5cm}+O((\alpha-1)^4)
\end{eqnarray}
Here we have chosen to expand around $\alpha=1$ since that corresponds to the case when two angular momenta are equal and one is zero. The first term in the expansion should hence be compared to the result in \cite{Beisert:2003xu}. It is also possible to compare the full result to the corrections for quantizing the three spin solution with two spins equal in \cite{Frolov:2003tu}.
We find that the above is exactly the expansion of 
\begin{equation}\label{result}
\gamma=\frac{\lambda\alpha}{2L}+\frac{\lambda}{L^2}p\sqrt{p^2+4-6\alpha-2\sqrt{4p^2-(4p^2+8)\alpha+9\alpha^2}}
\end{equation}
around $\alpha=1$. This coincides with the result in \cite{Frolov:2003tu} after we have divided the second term in (\ref{result}) by two, since (\ref{result}) is the contribution from two fluctuations.
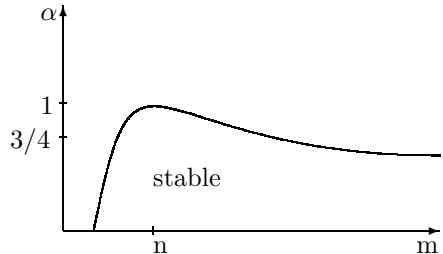
\begin{figure}
\hspace{1cm}
\begin{picture}(5,5)
\put(0,0){\vector(1,0){5}}
\put(0,0){\vector(0,1){3}}
\cbezier(0.4,0)(1,3)(1,1)(5,1)
\put(-0.3,2.8){$\alpha$}
\put(4.7,-0.3){m}
\put(-0.05,1.7){\line(1,0){0.1}}
\put(-0.3,1.55){1}
\put(-0.05,1.25){\line(1,0){0.1}}
\put(-0.7,1.1){3/4}
\put(1.2,-0.05){\line(0,1){0.1}}
\put(1.2,-0.3){n}
\put(1.2,0.6){stable}
\end{picture}
\caption{\label{figure4}The stability of the solution will depend on the winding number.}
\vspace{-0.3cm}
\end{figure}

Studying the expression in (\ref{result}) we find that for certain values of the parameter it becomes imaginary. This we interpret as an instability. The spin chain is an Hamiltonian system and should not have complex energy states. For the Bethe ansatz, the classical instability is manifested by a root moving off the real axis. A well defined quantum state require a complex Bethe root to appear with its conjugate. So an unstable mode by itself does not correspond to a well defined quantum state.
To study this more closely let us consider the case when the fluctuation is massless, that is when moving roots does not give any contribution to the energy/anomalous dimension. This is obtained putting the second term in (\ref{result}) to zero. The moved roots give zero contribution when
\begin{equation}
\alpha=1-\frac{p^2}{4}.
\end{equation}
According to the definition of $\alpha$ it has to be somewhere between zero and one. This means that we can have massless modes only for $p=0$,$\pm1$ and $\pm2$. $p=\pm1$ corresponds to $\alpha=3/4$ and the massless state found in \cite{Frolov:2003tu}. The case $p=\pm2$ is special, the state is massless for any value of $\alpha$. From the definition of $p$ we have that $n=m(p+1)$, $n$ is positive which means that $p$ must be greater or equal to zero if $m=1$. We have also assumed that $m\neq n$, which means that $p=0$ is not allowed. 
If $p=1$, then the mode is massless for $\alpha=3/4$, which correspond to $J'\leq3/2J$. Above this value, $\gamma$ is complex, so the mode is tachyonic. Below this value the mode is massive.

From the value of $\mu$ given in (\ref{positions}) the roots are always on the real axis if $p>2$. If $p<2$ then the roots are on the real axis if $\alpha<1-p^2/4$, they are at the origin if $\alpha=1-p^2/4$ and they are on the imaginary axis if $\alpha>1-p^2/4$. 
Since the positions of the roots are given for any $m$ it is possible to draw conclusions about the stability for general winding numbers of the corresponding string solutions.  This is shown in figure \ref{figure4}.

\section{Conclusions}
In conlusion we have considered the operators dual to circular strings rotating in $S^5$. We have found their scaling dimensions up to corrections of order $1/L^3$. This corresponds to the energy of the spinning strings when quantum corrections are taken into account. We have introduced fluctuations to the solution on the gauge side by moving Bethe roots around in the complex plane, being careful to preserve the representation. The result is an expression for the anomalous dimension that exactly agrees with the energy of the quantized strings found earlier.

We have also obtained that an instability on the string side corresponds to roots being moved off the real axis. As $\alpha$ is increased the root corresponding to a fluctuation will move from the real axis to the origin, where it becomes massless, and then off on the imaginary axis. As it moves onto the imaginary axis the instability occurs.

\vspace{0.3cm}
In the course of this work \cite{Hernandez:2004uw} appeared where the non-linear sigma model describing the continuum limit of the $SU(3)$ spin chain was constructed. How to relate the gauge theory computations using spin chains to these continuous models are described in \cite{Kruczenski:2003gt,Kruczenski:2004kw,Stefanski:2004cw}. It turns out that in this framework it is also possible to find the spectrum from fluctuations. 

\subsubsection*{Acknowledgements}
I would like to thank Joseph Minahan for many helpful discussions, for introducing me to the problem and reading the manuscript. I would also like to thank  Johan Engquist, Mikael Smedb\"{a}ck and Konstantin Zarembo for interesting discussions.

\bibliographystyle{unsrt}
\bibliography{Betheref}
\end{document}